\documentclass[reprint,aps,pre,floatfix]{revtex4-2}
\usepackage[utf8]{inputenc}
\usepackage{amsmath}
\usepackage{graphicx}
\usepackage{physics}
\usepackage[citecolor=black,%
filecolor=black,%
linkcolor=blue,%
urlcolor=black,%
colorlinks=true]{hyperref}
\usepackage{lineno}
\usepackage{xcolor}

\usepackage{natbib}

\newcommand{\upd}{\text{d}} 

\newcommand{\slfrac}[2]{\left.#1\middle/#2\right.}
\renewcommand{\vec}{\boldsymbol}

\usepackage{natbib}
\usepackage{lineno}

\begin{document}

%\preprint{APS/123-QED}
%\linenumbers
\title{Spatio-temporal patterns of active epigenetic turnover} 

\author{Fabrizio Olmeda$^{1,2,3}$, Misha Gupta$^{1,4}$, Onurcan Bektas$^{2}$ and Steffen Rulands$^{1,2,*}$
} 
\affiliation{$^{1}$Max-Planck-Institute for the Physics of Complex Systems, Noethnitzer Str. 38, 01187 Dresden, Germany \\
$^{2}$Ludwig-Maximilians-Universit\"at M\"unchen, Arnold-Sommerfeld-Center for Theoretical Physics, Theresienstr. 37, 80333 M\"unchen, Germany\\
$^3$Institute of Science and Technology Austria, Am Campus 1, 3400 Klosterneuberg, Austria \\
$^4$Department of Organismic and Evolutionary Biology, Harvard University, Cambridge, MA 02138, USA
}

\begin{abstract}
DNA methylation is a primary layer of epigenetic modification that plays a pivotal role in the regulation of development, aging, and cancer. The concurrent activity of opposing enzymes that mediate DNA methylation and demethylation gives rise to a biochemical cycle and active turnover of DNA methylation. While the ensuing biochemical oscillations have been implicated in the regulation of cell differentiation, their functional role and spatio-temporal dynamics are, however, unknown. In this work, we demonstrate that chromatin-mediated coupling between these local biochemical cycles can lead to the emergence of phase-locked domains, regions of locally synchronized turnover activity, whose coarsening is arrested by genomic heterogeneity. We introduce a minimal model based on stochastic oscillators with constrained long-range and non-reciprocal interactions, shaped by the local chromatin organization. Through a combination of analytical theory and stochastic simulations, we predict both the degree of synchronization and the typical size of emergent phase-locked domains. We qualitatively test these predictions using single-cell sequencing data. Our results show that DNA methylation turnover exhibits surprisingly rich spatio-temporal patterns which may be used by cells to control cell differentiation.
%These dynamics suggest that epigenetic regulation is not only context-dependent but may also be actively modulated over time to influence developmental trajectories. 
\end{abstract}

\maketitle
%\section{Introduction}
Epigenetics collectively describes biomolecular processes that regulate cell behavior beyond what is encoded in the DNA sequence. These processes involve the folding of the DNA and chemical modifications of the DNA and histone tails. DNA methylation primary affects cytosines next to guanines, termed CpG pairs. The positioning of DNA methylation marks in the genome has been shown to be relevant for the assignment of cell types during embryonic development~\cite{parry2021active,agostinho2023epigenetic}, it is closely associated with aging~\cite{hernando2019ageing}, and alterations in related enzymes are one of the hallmarks of blood cancer~\cite{egger2004epigenetics}.

\begin{figure}[tb]
\includegraphics[width=0.5\textwidth]{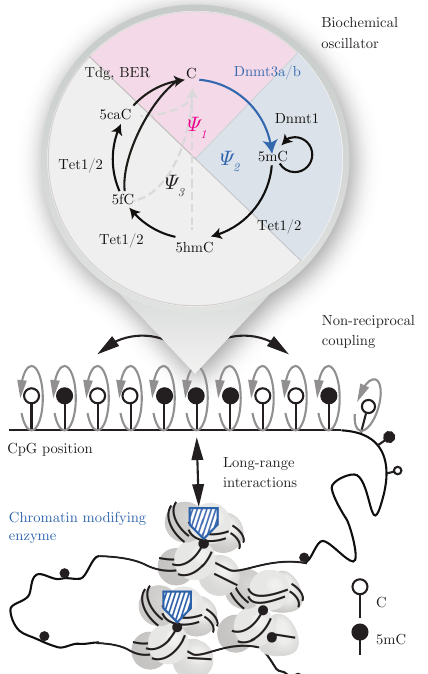}
\caption{ The biochemistry of DNA methylation turnover consists of  a biochemical cycle.  The DNA methylation cycle is divided into three phase intervals, $\psi_1,\psi_2,\psi_3$. CpGs in the first phase interact with restricted  long-range interactions with CpGs that are in the second phase.}
\label{fig:Figure1}
\end{figure}

DNA methylation marks are established by a family of DNMT3 enzymes and are maintained during DNA replication by DNMT1 enzymes, among others. DNA methylation marks can also be actively removed through a chain of biochemical reactions involving TET enzymes. TET enzymes modify the methylated cytosines (5mC) by oxidating 5mC to hydroxymethyl-cytosine (5hmC) then to  formyl-cytosine (5fC), and finally to carboxyl-cytosine (5caC) (\hyperref[fig:Figure1]{Fig.~\ref{fig:Figure1}}) \cite{parry2023dynamic,he2011tet,parry2021active}. 
Paradoxically, in several biologically relevant contexts, cells express antagonistic enzymes that methylate and demethylate the DNA, namely members of the protein families DNMT3 and TET \cite{szulwach20115, parry2021active}. These contexts include the priming of pluripotent cells for differentiation in early development~\cite{rulands2018genome,Parry2023.01.11.523441} and the differentiation of hematopoetic stem cells~\cite{cimmino2011tet}. This coexpression leads to a biochemical, largely irreversible cycle, in which a CpG undergoes cyclic chemical turnover involving the conversion of cytosines (C) to methylated cytosines (5mC), then to hydroxymethylated cytosines (5hmC) and via multiple intermediary steps back to the unmodified cytosine. The periodicity of this oscillation is approximately two hours, and the oscillation phase is coupled genome-wide to a limited degree \cite{rulands2018genome}. Oscillatory DNA methylation turnover has been specifically linked to enhancer regions~\cite{rulands2018genome,parry2023dynamic} and to inducible genes targeted by the estrogen receptor $\alpha$~\cite{kangaspeska2008transient,metivier2008cyclical}. Active turnover of DNA methylation seems to be necessary for the regulation of cell differentiation in development~\cite{Parry2023.01.11.523441}.  

Experimental work on active DNA methylation turnover has mainly focused on DNA sequencing technologies such as bisulfite sequencing ~\cite{booth2012quantitative}. State-of-the-art technologies cover around one in five CpGs in a given cell and cannot distinguish between more than two chemical modifications. Methods that can distinguish between C, 5mC and 5hmC in single cells have orders of magnitude lower resolution~\cite{experimentaldatabai}. Because sequencing also relies on destructive sampling of the cells and DNA, dynamic information about DNA methylation turnover can only be inferred indirectly. Therefore, despite active turnover of DNA methylation being associated with relevant biological scenarios, its spatio-temporal dynamics and biological function are unknown. Theoretical predictions are hence pivotal for understanding the spatio-temporal dynamics of DNA methylation turnover and inform targeted experimental approaches.

Here, we investigate the range of spatio-temporal behaviors that DNA methylation turnover can exhibit. We map the interplay between DNA methylation turnover and the geometrical configuration of the chromatin to a driven oscillator model with non-reciprocal and non-local interactions. We show that DNA methylation turnover leads to phase-locked domains that are globally weakly coupled. Using analytical and numerical calculations, we demonstrate that the degree of global synchronization increases with the coupling strength and saturates at a partially synchronized regime in the limit of infinite coupling strengths. Phase-locked domains coarsen over time much larger than the characteristic time scale of active epigenetic turnover for weak disorder in the frequencies of individual oscillators. We further show that genomic disorder stemming from the DNA sequence enhances local synchronization and leads to an arrest of coarsening and constant size of phase-locked domains. Finally, we use sequencing data that simultaneously identify C, 5mC and 5hmC states to qualitatively test key predictions of our theory.

\section*{Results}
%\subsection*{Derivation of phase oscillators dynamics with restricted long-range interactions}
We first derive a stochastic lattice model of DNA methylation turnover to investigate the potential spatio-temporal patterns emerging from these interactions. To this end, we consider a lattice of $N$ sites. The state in the DNA methylation turnover cycle at each position, $i$, is described by a phase variable $\phi_i$, which can take values in $\{1,\ldots,\Omega\}$, \hyperref[fig:Figure1]{Fig.~\ref{fig:Figure1}}.  The state of the system is then described by a vector containing the phase values at each site, $\vec{\phi}=(\phi_1,\ldots,\phi_N)$. The phase at site $i$ is cyclically advanced by 1 with a rate $\tilde{\omega}_i =\omega_i+k_i(\vec{\phi})$. This rate has a site-dependent, but constant, contribution $\omega_i$, denoting the intrinsic rate of chemical conversions at that position $i$. The function $k_i(\vec{\phi})$ captures the contribution of the phases at other sites. While interactions are local in physical space, the folding of the DNA and chromatin can induce long-range interactions along the one-dimensional DNA sequence. This folding is reflected in enzyme binding rates, as the establishment of DNA methylation marks is associated with the local compaction of chromatin~\cite{olmeda2021inference,lovkvist2016dna,li2022dna}. We describe the local structure of chromatin in terms of an exponent $\gamma$ describing the inner contact probability, such that $k_i(\vec{\phi})$ decays like $|i-j|^{-\lambda}$. This exponent $\gamma$ takes values between 2.1 for a self-avoiding random walk and 1/3 for space-filling chromatin. On the nano scale, chromatin is organized into heterogeneous structures (``clutches") \cite{otterstrom2019super}. On the associated length scale, we therefore assume that long-range contacts are exponentially suppressed and $k_i(\vec{\phi})$ is exponentially cutoff  \cite{olmeda_field_2024} . This cutoff also ensures the existence of the thermodynamic limit.

The local compaction of chromatin influences the chemical conversion rates in previous steps of the cycle, such that interactions are non-reciprocal. For example, methylated sites recruit more DNMT3 enzymes and thereby increase conversion rates of unmethylated sites in the vicinity \cite{olmeda2021inference}. The reverse is generally not true. To consider this mathematically,  we denote the set of chemical states that are associated with compacted chromatin by $\psi_2$, biologically associated with the binding of DNMT3 enzymes and the methylation of CpGs. Chemical states in $\psi_2$ induce an increase in the rate of state transitions in a set of states preceding $\psi_2$. We denote these preceding states by $\psi_1$ (\hyperref[fig:Figure1]{Fig.~\ref{fig:Figure1}}). In the mean-field limit this then leads to a contribution to the transition rate of the form \cite{olmeda2021inference,hinrichsen2006non,ginelli2006contact},
\begin{equation}
\label{eq:KernelDefinition}
k_{i}\left(\vec{\phi}\right) = J \sum_{k=1,k \neq i }^N \mathcal{I}_{\phi_i,\psi_1} \mathcal{I}_{\phi_k,\psi_2}\frac{ e^{-m|k -i|}}{|k-i|^{\lambda}}\, ,
\end{equation}
where $J$ denotes the coupling strength between sites. The indicator function $\mathcal{I}_{A,B}$ is 1 if $A\in B$ and 0 otherwise. $m = N^{-1}\sum_{j=1}^N \mathcal{I}_{\phi_j,\psi_2}$ is then the fraction of sites in phase interval $\psi_2$. The DNA methylation dependence of the exponential cutoff in Eq.~\eqref{eq:KernelDefinition}  is a consequence of screening due to a feedback between DNA methylation and chromatin \cite{olmeda2021inference,olmeda_field_2024}.

The time evolution of the joint probability of finding a given lattice configuration $\vec{\phi}$ at time $t$, $P(\vec{\phi},t)$, follows a master equation of the form
\begin{equation}
\label{eq:MasterEquationClock}
    \pdv{P(\vec{\phi},t)}{t} = \sum_{i=1}^N (\mathrm{E}_i^{-1}-1)\left[\omega_i + k_{i}\left(\vec{\phi}\right)\right]P(\vec{\phi},t)\, .
    %k_{i}\left(\vec{\phi}\right) &= J \sum_{k=1/k \neq i }^N \delta_{\phi_i,\psi_1} \delta_{\phi_k,\psi_2}\frac{ e^{-m|k -i|}}{|k-i|^{\lambda}}\, ,
\end{equation}
Here, we made use of the the shifting operator $\mathrm{E}_i$ which raises or lowers the phase variable in a function $G$ on a given site by one, $\mathrm{E}^{\pm 1}_i G(\vec{\phi}) = G(\phi_i,\ldots, \phi_i\pm 1,\ldots, \phi_N)$ \cite{Kampen}. 

\subsection*{A continuum theory of DNA methylation turnover}
We now ask under which conditions DNA methylation turnover can synchronize and characterize potentially emerging spatio-temporal structures. To gain an initial understanding, we first neglect the disorder that results from genomic variations in CpG positions. To begin, we separate the phase variable $\phi_i$ into a deterministic part, which scales as the numbers of states in the cycle, $\Omega$, and a stochastic component, which scales as the square root of $\Omega$ (system size expansion)~\cite{Kampen,Jorg2017}. In doing so, we promote $\phi_i$ to be a continuous variable that satisfies $\phi_i = \Omega \Phi_i(t) + \Omega^{1/2}\xi_i(t)$.

After substituting this expression into the master equation and collecting terms of equal order in $\Omega$, \hyperref[sec:DiscretePhaseExpansion]{Appendix~\ref{sec:DiscretePhaseExpansion}}, we obtain a Langevin equation describing the time evolution of the phase at each site,
\begin{equation}
\label{eq:langevinOscillators}
    \dv{\phi_i}{t} = w_{i} + f_1(\phi_i)\sum _{k=1,k \neq i}^{N}\frac{J e^{{{-m|k -i|}}}}{|k-i|^{\lambda}}f_2(\phi_k) + \sqrt{2\omega_i}\xi_i(t)\, .
\end{equation}
Here, $\xi_i(t)$ is Gaussian white noise with zero mean and unitary variance, $\langle \xi_i(t)  \xi_j(t') \rangle = \delta(t-t')\delta_{i,j}$ . The functions $f_{1,2}(\phi)$ are the continuous versions of $\mathcal{I}_{\phi,\psi_{1,2}}$ , respectively.
Eq.~\eqref{eq:langevinOscillators} has the form of a stochastic Kuramoto model \cite{Kuramoto2003,Gupta2014} with non-reciprocal,  restricted long-range interactions. 
In the following section, we will derive an exact, coarse-grained description of Eq.~\eqref{eq:langevinOscillators} in order to rigorously investigate the emergence of locally synchronized states.

\begin{figure*}
 \begin{center}
\includegraphics[width=0.99\textwidth]{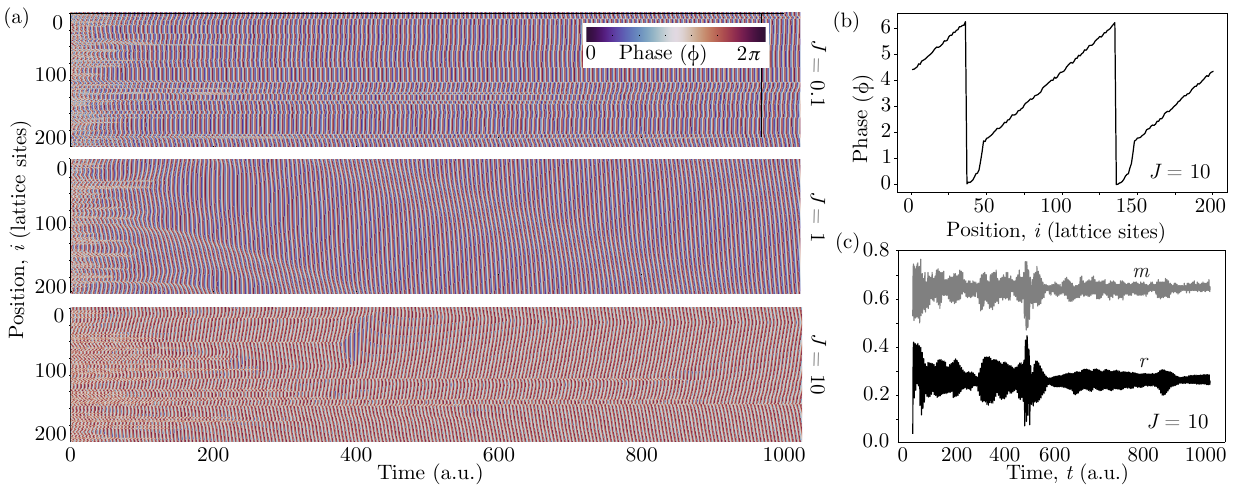}
\caption{(a) Heatmaps showing the spatio-temporal evolution of the deterministic part of Eq.~\eqref{eq:langevinOscillators} starting from random initial conditions and identical intrinsic frequencies for all sites, i.e. $\omega_i = 1$ for all $i$. (b) Spatial distribution of phases after $10^6$ time steps, highlighting the phase-locked structures for high-coupling strength ($J = 10$). (c)  The Kuramoto order parameter $r$ and average DNA methylation $m$ exhibit fluctuations and oscillations before approaching a steady state. Details of the numerical simulations are given in \hyperref[sec:NumericalMethods]{Appendix~\ref{sec:NumericalMethods}}.}
\label{fig:ExampleSyncronization}
\end{center}
\end{figure*}

A suitable order parameter to quantify the degree of system-wide synchronization is  the Kuramoto order parameter $r(t)$ ~\cite{Kuramoto2003}, defined as
\begin{equation}
    r(t)e^{i\psi(t)}= \frac{1}{N}\sum_{i=1}^{N} e^{i\phi_i(t)} \, .
\end{equation}
We further define a function, $\rho(\theta,\omega,z,t)$ as the density of oscillators with phase $\theta$, position $z$ and frequency $\omega$ at time $t$. The time evolution of $\rho$ follows a continuity equation derived by coarse-graining Eq.~\eqref{eq:langevinOscillators} (\hyperref[sec:FPDerivation]{Appendix~\ref{sec:FPDerivation}}) of the  phase $\phi_i$,
\begin{equation}
\label{eq:continuity}
    \frac{\partial \rho (\theta,\omega,z)}{\partial t} = -\frac{\partial}{\partial \theta}\left[(\omega+\tilde{\nu}) \rho (\theta,\omega,z,t)\right] \, . 
\end{equation}
This equation is a transport equation and it describes the translation of the density $\rho$ along the phase cycle with a velocity given by constant term $\omega$ and an additional contribution stemming from long-range interactions, which reads
\begin{equation}
\label{eq:velocity_rho}
     \tilde{\nu} = J f_1(\theta)\left[  \int \upd\omega \upd{z'} \upd{\theta'} \, \frac{ g(\omega')e^{-m {z'}}}{|{z'}|^{\lambda}}  f_2({\theta'}) \rho({\theta'},\omega',z-{z'})\right]\,. 
\end{equation}
For a detailed derivation see \hyperref[sec:FPDerivation]{Appendix~\ref{sec:FPDerivation}}. The function $g(\omega)$ denotes the distribution of intrinsic frequencies of DNA methylation turnover cycles. It is determined by enzyme binding and unbinding rates, as well as by the availability of metabolites and chemical reaction products necessary for the conversion between different chemical states of the cytosine. Spectral analysis of a sequencing time course revealed that oscillation frequencies are region-specific but constrained to an overall period between 1 and 2 hours~\cite{rulands2018genome}.

The phase cycle velocity depends only on the global fraction of sites that carry DNA methylation marks (sites in the phase interval $\psi_2$), $m$ and the order parameter, $r$, which are defined as,
\begin{equation}
      \label{eq:definition_meth}
 \begin{split}
    m &= \int  \upd \theta \upd \omega \upd z\, g(\omega) f_2(\theta')\rho(\theta,\omega,z)\,, \\
    re^{i\psi} &= \int  \upd\theta \upd\omega \upd z  \, \rho(\theta,\omega,z)  g(\omega)e^{i\theta}\, ,
 \end{split}
  \end{equation}
respectively. The integral over $z$ goes from zero to one. Together, equations ~\eqref{eq:definition_meth},  ~\eqref{eq:velocity_rho} and ~\eqref{eq:continuity}  define the density $\rho$ in the continuous limit.

\subsection*{Coarsening of phase-locked domains}
We numerically solved the deterministic part of Eq.~\eqref{eq:langevinOscillators}, 
 to quantitatively understand how far individual oscillators synchronize due to long-range interactions, \hyperref[fig:ExampleSyncronization]{Fig.~\ref{fig:ExampleSyncronization}a}. These simulations show that DNA methylation turnover at short times synchronizes in finite-size domains. Within these domains, phase differences between neighboring sites remain constant in time (phase-locking) and biochemical oscillators are weakly synchronized system-wide. These domains slowly coarsen over time, such that, at late times, the system achieves system-wide phase-locking, \hyperref[fig:ExampleSyncronization]{Fig.~\ref{fig:ExampleSyncronization}b}. 
The average DNA methylation $m$ and the Kuramoto order parameter $r$ oscillate around constant values in the long-term limit, \hyperref[fig:ExampleSyncronization]{Fig.~\ref{fig:ExampleSyncronization}c}. 
The coarsening dynamics of phase-locked domains depend non-monotonically on the coupling strength. For small values of the coupling, $J = 0.1$, neighboring phase-locked domains coarsen slowly. However, for larger values of the coupling, $J = 1$, coarsening occurs faster. Interestingly, for $J = 10$, once larger domains are formed, their coarsening slows down again compared to the intermediate coupling strength, $J = 1$. These observations were consistent across 1000 simulations with randomly sampled initial conditions.
In the next section, we derive the effect of restricted long-range interactions in the partial synchronization of DNA methylation domains. We will then develop a strong-disorder renormalization method to quantify the coarsening time and its dependence on the interaction strength $J$.

%\subsection*{Synchronization and formation of spatial phase-locked domains}
\begin{figure}
    \centering
\includegraphics[width=0.45\textwidth]{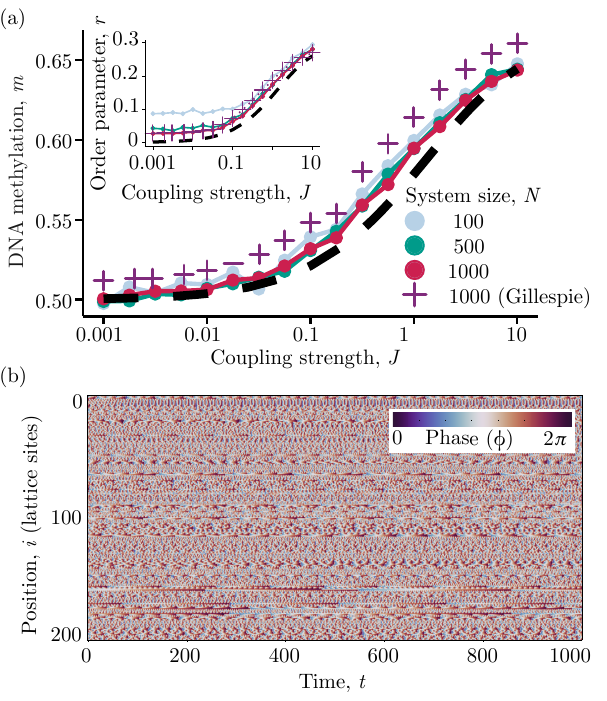}
    \caption{ (\textit{a}) Kuramoto order parameter $\langle r \rangle$ and $\langle m \rangle$ (inset) as a function of the interactions strength $J$. The average is taken over 100 realizations of deterministic (points) and stochastic (crosses) numerical simulations of the model in Eq.~\eqref{eq:langevinOscillators}. Different colors represent different values of the lattice size, $N$. Theoretical predictions from Eqs. \eqref{eq:rho2Final} and \eqref{eq:continuity} are shown as a black dashed line. For the stochastic simulations, we set the number of states to 100. Here, the frequencies $\omega_i$ are random numbers drawn from an exponential distribution with unitary mean. (\textit{b}) Long-term behavior of oscillators with exponential distribution of frequencies  with intermediate-coupling strength, $J=1$. Details of the numerical simulations are given in \hyperref[sec:NumericalMethods]{Appendix~\ref{sec:NumericalMethods}}.}
\label{fig:ComparisonKuramotoLongrange}
\end{figure}

We investigate how disorder in the interactions, given by the position of CpGs base pairs, by computing the spatially homogeneous stationary solution of Eq.~\eqref{eq:continuity} (\hyperref[sec:StatSol]{Appendix~\ref{sec:StatSol}}). We find that there are two possible solutions: First, oscillators may be phase-locked at identical frequencies such that $\omega + \tilde{\nu} =0 $. Secondly, oscillators may rotate with frequencies given by  $(\omega + \tilde{\nu}) \rho = C(\omega)$, where $C(\omega)$ is a constant, which is determined by the normalization of the density $\rho$ of oscillators \cite{Acebron2005}. Solving for both branches of solutions \hyperref[sec:Mvalue]{Appendix~\ref{sec:Mvalue}}, the fractions of sites carrying DNA methylation marks follows,
\begin{equation}
\label{eq:rho2temp}
    m =   \int_0^{\infty} \upd\omega  \frac{  2(A(\lambda) J+\omega)}{3A(\lambda)
   J+4 \omega }  g(\omega) \, .
\end{equation}
where $A(\lambda) =  m^{\lambda} \Gamma(1-\lambda,0,m )$.

For an exponential distribution of intrinsic frequencies, $g(\omega)=\mu^{-1}\exp(\mu\omega)$, $m$ then satisfies  the  self-consistent equation,
\begin{equation}
\label{eq:rho2Final}
 m = \frac{1}{2} -   \frac{\mu JA(\lambda)}{8} e^{\frac{3 \mu J A(\lambda)}{4}} \text{Ei}\left(-\frac{3 \mu J A(\lambda)}{4}  \right)\, ,\\
\end{equation}
where $\text{Ei}(x) = -\int_{-x}^{\infty} \text{d}y \, e^{-y}/y$ is the exponential integral function. In the limits of strong and weak interactions, the integrals can be evaluated straightforwardly: for $J\rightarrow 0$, we obtain $m = {1}/{2}$ and for $J\rightarrow \infty$, we obtain $m ={2}/{3}$. The results in these limits can be obtained by straightforward evaluation of the integrals: in the case of $J\to 0$, the expression for $m$ simplifies to $m = \int \text{d}\theta f_2(\theta)/2\pi$, which yields with our choice of $f_2$ a value of $1/2$. 
If $J \rightarrow \infty$, oscillators that are in the phase interval $\psi_1$, which is the only one affected by interactions, spend an infinitesimally small amount of time in that phase interval, such that the density of oscillators is effectively restricted in the phase domain $[\pi/2,2\pi]$. The oscillators in this interval do not interact. Here, on average, $m = \int \text{d}\theta f_2(\theta)/(3\pi/2)$, which for our specific choice of $f_2$  gives the value $2/3$.

For intermediary values of $J$, Eq.~\eqref{eq:rho2Final} cannot be further simplified and must be solved numerically.  \hyperref[fig:ComparisonKuramotoLongrange]{Figure~\ref{fig:ComparisonKuramotoLongrange}a} shows the analytical prediction, Eq.~\eqref{eq:rho2temp}, compared to stochastic and mean-field lattice simulations for different lattice sizes (\hyperref[sec:DiscretePhaseExpansion]{Appendix~\ref{sec:DiscretePhaseExpansion}}). For the stochastic simulations, we varied the number of phase states between 10 and 100 with consistent results. We attribute the small discrepancy between numerical simulations and theoretical results (\hyperref[fig:ComparisonKuramotoLongrange]{Fig.~\ref{fig:ComparisonKuramotoLongrange}a} inset) to finite-size effects and the formation of spatial structures. The Kuramoto order parameter, $r$, can be computed once the value of $A(\lambda)$ is known numerically from Eq.~\eqref{eq:continuity}. Specifically, because $m$ saturates at a constant value in the long-term limit, $r$ must be strictly smaller than $1$, and full synchronization cannot be achieved in the limit $N\to\infty$.  \hyperref[fig:ComparisonKuramotoLongrange]{Figure~\ref{fig:ComparisonKuramotoLongrange}b}  shows typical spatial distribution of phases with an exponential distribution of frequencies. Unlike the case with equal intrinsic frequencies for all oscillators, heterogeneity in frequencies prevents the formation of phase-locked domains. However, in the next section, we will show that genomic disorder leads to domain formation for oscillators with heterogeneous intrinsic frequencies. Taken together, our calculations and simulations reveal that DNA methylation turnover does not exhibit a phase transition from an asynchronous to a synchronous state, which is typical of Kuramoto-like models \cite{Acebron2005}. Rather, we find that synchronization and DNA methylation increase continuously with the interaction strength. However, for all coupling strengths $J>0$ the domains synchronize partially and these local spatial structures affect the synchronization and domain coarsening. 

\begin{figure*}
   \centering
\includegraphics[width=0.99\textwidth]{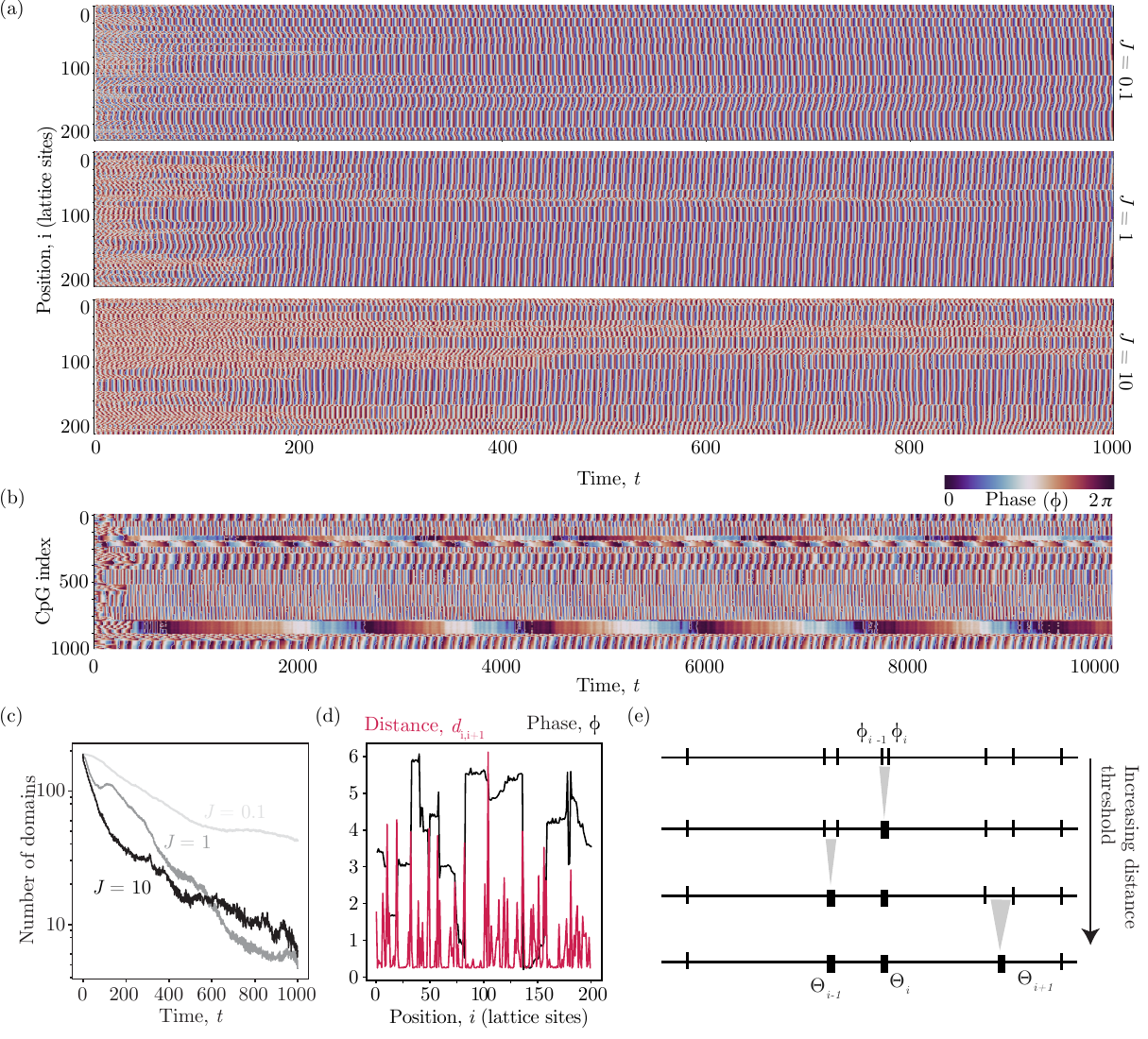}
   \caption{(\textit{a}) Numerical simulations the deterministic part of Eq.~\eqref{eq:langevinOscillators} starting from random initial conditions and a delta distribution of intrinsic frequencies $\omega_i = 1$ for all $i$. The distances between neighboring sites are sampled from the first 200 CpGs of Chromosome 1 of the mouse genome. We tested different regions of the genome and we found consistency in the results. (\textit{b}) At long times, the phase-locked domain structure is independent of the coupling strength. (\textit{c}) Number of of phase-locked domains over time for different coupling strengths, $J = 0.1,1,10$ for simulations with $1000$ sites. Details on the inference of domain numbers are given in \hyperref[sec:NumericalMethods]{Appendix.~\ref{sec:NumericalMethods}}. (\textit{d}) Comparison of the distance to the nearest CPG and phase variable $\phi$. Strong changes in the density of CpGs are associated with changes in the phase variable. Data was taken at $1e4$ oscillation periods. (\textit{e}) Schematic illustrating the strong-disorder renormalization group scheme. At each step, we merge the closest neighboring blocks. We repeat this process until the distances between the chosen adjacent blocks is larger than the average distance between CpGs.}
    \label{fig:Fig4}
\end{figure*}

\subsection*{Effect of genomic disorder on phase locking}
In the genome, the positions of CpGs are heterogeneously distributed, with many of them clustered in so-called CpG islands. Because the effective strength of interactions decays with the genomic distance, Eq.~\eqref{eq:KernelDefinition}, the coupling strength varies randomly with the genomic position $i$ (quenched disorder). To investigate the role of such disorder on the degree of phase locking, we performed numerical simulations of Eq.~\eqref{eq:langevinOscillators}, with distances between sites equal to those of different regions of chromosome 1 of the mouse genome, \hyperref[fig:Fig4]{Fig.~\ref{fig:Fig4}a,b}. 
We rescaled distances so that the average distance between neighboring CpGs is equal to one. With genomic disorder, the coarsening dynamics of phase-locked domains is slower, \hyperref[fig:Fig4]{Figure~\ref{fig:Fig4}a},  compared to the regular, one-dimensional lattice, \hyperref[fig:ExampleSyncronization]{Fig.~\ref{fig:ExampleSyncronization}a}. After approximately 1e4 oscillation periods, \hyperref[fig:Fig4]{Fig.~\ref{fig:Fig4}c}, the coarsening is arrested, resulting in stable, phase-locked structures, \hyperref[fig:Fig4]{Fig.~\ref{fig:Fig4}b}. In the steady state, domain boundaries are determined by the distribution of distances between neighboring CpGs: they tend to coincide with genomic regions in which the density of CpGs changes strongly, \hyperref[fig:Fig4]{Fig.~\ref{fig:Fig4}d}.  

To investigate the coarsening dynamics of phase-locked domains, we now employ \textit{strong disorder renormalization} of Eq.~\eqref{eq:langevinOscillators} \cite{PhysRevE.80.046210,PhysRevE.80.036206}, \hyperref[fig:Fig4]{Figure~\ref{fig:Fig4}e}. 
This iterative renormalization procedure defines a coarse-graining scheme on blocks of neighboring CpGs. Initially, each block contains exactly one CpG. At each renormalization step, we select the pair of neighboring blocks with the closest genomic distance, and hence the strongest coupling strength. We then define a new block that incorporates the pair of chosen CpGs. We defined the phase of the new block, $\Theta_i$,  as the average phase of the coarse-grained blocks, weighted by the number of CpGs. The size of the coarse-grained block , $N_i$, given by the sum of the number of CpGs of the blocks that were averaged, $N_i=n_i+n_{i+1}$, and its intrinsic frequency by the average weighted by the number of CpGs in each block, $\Omega_i = \slfrac{(n_i \omega_i + n_{i+1} \omega_{i+1})}{(n_i + n_{i+1})}$. We stop the renormalization procedure when the typical distance between blocks is greater than the average distance between CpG sites. After all these steps, Eq.~\eqref{eq:langevinOscillators} is effectively described by a model with interactions between neighboring blocks,
\begin{equation}
N_i \pdv{\Theta_{i}}{t} = N_i \Omega_i + Je^{-m}f_1(\Theta_i) \left[f_2(\Theta_{i-1}) + f_2(\Theta_{i+1})\right]\, .
\end{equation}
To estimate the time of coarsening of phase-locked domains we compute the dynamics of the phase difference between two adjacent coarse-grained blocks, $\Theta_i - \Theta_{i-1}$. If all oscillators have identical intrinsic frequencies this coarse-grained phase difference decreases exponentially with a typical time scale given by
\begin{equation}
 \label{eq:rg_genome}
\tau_i \approx \frac{N_i}{Je^{-m}} \left( \frac{1}{N_i} + \frac{1}{N_{i-1}} \right)^{-1} \,  .
\end{equation}
This result shows that the coarsening-time scales quadratically with the typical number of CpGs in a coarse-grained block. 
In particular, localized regions, such as CpG islands, which have  approximately 100 CpGs, will be phase-locked for approximately $1e4$ oscillations for a value of $Je^{-m}$ of order one. This result is confirmed by our simulations (\hyperref[fig:Fig4]{Fig.~\ref{fig:Fig4}}) for even smaller domains. For a typical oscillation period of 2 hours, this translates to at least 27 months, which is longer than the typical lifetime of the model system that was used to experimentally determine the periodicity (mouse). Therefore, we expect that arrested phase-locked domains are stable across all biologically relevant time scales.

\begin{figure*}[ht]
    \centering
\includegraphics[width=0.99\textwidth]{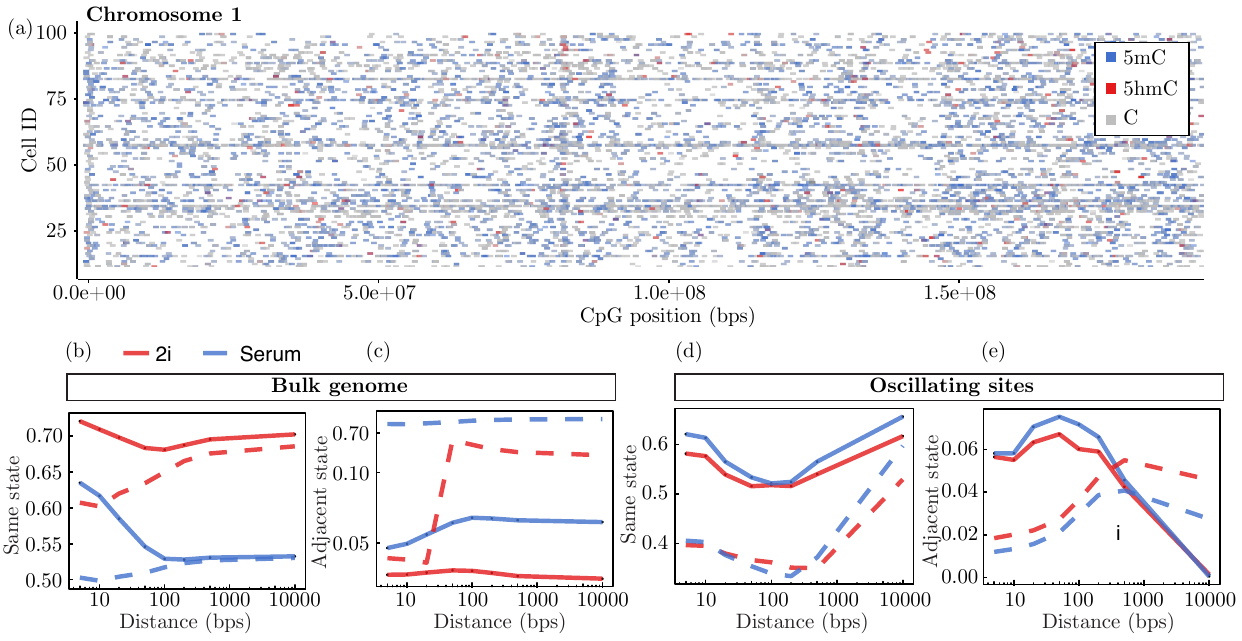}
   \caption{(\textit{a}) Representative example of experimental sequencing data, with individual CpG sites colored according to their measured state in the biochemical cycle. (\textit{b}) Emprical probability of finding two consecutive CpGs with the same state in the biochemical cycle (solid line) compared to the case of statistical independence (dashed line). The distance dependence of the null hypothesis stems from the non-uniform distribution of CpG sites. (\textit{c}) Empirical probability of finding neighboring CpG states in adjacent states of the biochemical cycle (solid line, in this case C and 5mC). Dashed lines depict the prediction under the null hypothesis of statistical independence. In (\textit{d}) and (\textit{e}) show respectively the same analysis as in (\textit{b}) and (\textit{c}) but filtered for sites that putatively undergo DNA methylation turnover.}
    \label{fig:Fig5}
\end{figure*}

Our theoretical work shows that the dynamics of DNA methylation turnover exhibits behavior that is not predicted by driven oscillator models studied in the theoretical literature~\cite{Acebron2005}. The Kuramoto model with long-range interactions~\cite{Gupta2014} exhibits global synchronization if the coupling strength exceeds a threshold value. Here, non-reciprocal interactions lead to the formation of phase-locked domains which coarsen over time. This leads to global phase locking but not to global synchronization. Driven oscillator models have also been studied with disorder in the couplings. The Kuramoto model with disordered interactions leads to partial synchronization if the couplings can take negative values \cite{bonilla1993glassy}. In the biochemistry of DNA methylation turnover, the coupling must be positive. The disorder in the coupling between oscillators induced by the DNA sequence qualitatively changes this behavior in the asymptotic limit: it leads to an arrest of coarsening such that in this case global phase locking is not achieved.

\subsection*{Validation of theoretical predictions using sequencing data}

The precise quantitative validation of the spatio-temporal patterns predicted by the theory is highly challenging due to the lack of resolution in time, space, and biochemical phase of current sequencing technologies. To qualitatively test the existence of phase-locked domains, we use data recently generated with the SIMPLE-seq method that can distinguish the states C, 5mC, and 5hmC with low coverage in single cells ~\cite{experimentaldatabai}. We performed the primary bioinformatics analysis comprising sequence alignment and quality control as described in \hyperref[sec:DataAnalysis]{Appendix ~\ref{sec:DataAnalysis}}.  We focussed our analysis on mouse embryonic stem cells (mESCs) in 2i and serum conditions. These cells co-express the Dnmt1 or Dnmt3 and TET genes and therefore fulfill the biochemical conditions for active turnover of DNA methylation. MESCs in serum conditions have been shown to exhibit oscillatory dynamics in DNA methylation~\cite{rulands2018genome} and globally in nascent transcription~\cite{SHAH2018363} with a similar oscillation period of 2 hours. \hyperref[fig:Fig5]{Figure~\ref{fig:Fig5}a} gives an overview over the spatial distribution of the three states measured in mouse embryonic stem cells. As has been previously reported, the 5hmC mark occurs with a lower probability compared to the C and 5mC states.

%To detect phase-locking behavior we first asked whether  CpGs in genomic proximity correlate in the stage of the DNA methylation cycle. To this end, we  To begin, we asked whether CpG site

%Cells are further clustered based on their culture conditions, \textit{serum} and \textit{2i}. Briefly, the two conditions differ by their average methylation. It is clear that, as expected by our theoretical analysis and in accordance with the literature~\cite{rulands2018genome}, an increase in the average methylation leads to more oscillations (serum).
%In \hyperref[fig:Fig5]{Figure~\ref{fig:Fig5}d} we show the average phase of blocks of hundred consecutive CpGs for both these conditions. Cells are ordered based on their average 5mC methylation. 5mC methylations plays a similar role to time as cell that are release from  \textit{2i} to \textit{serum} gain methylation over time \cite{olmeda2021inference}.

To test whether DNA methylation turnover leads to phase-locked domains, we aimed to detect two signatures of phase-locked domains: first, we expect that CpGs in close vicinity show correlations in their DNA methylation state. Secondly, we expect that CpGs in close vicinity also exhibit a higher than expected tendency to be in adjacent states of the biochemical cycle. To begin, we first asked whether  CpGs in genomic proximity correlate in the stage of the DNA methylation cycle. To this end, we calculated the probability of finding two neighboring CpGs in the same DNA methylation state as a function of their distance. We compared this probability to the null hypothesis of statistical independence. Specifically, we resampled CpG states from a multinomial distribution using probabilities estimated as the observed fractions of each state across the data set. \hyperref[fig:Fig5]{Figure~\ref{fig:Fig5}b} shows that the probability of pairs of CpGs having the same DNA methylation mark is consistently higher than expected from statistical independence. This shows that the 5mC, 5hmC and C states are locally correlated, as has been reported before in the cases of C and 5mC.

We also calculated the probability that neighboring CpGs are in consecutive stages of the biochemical cycle. Specifically, we computed the probability of finding pairs of CpGs in adjacent states of the cycles (5mC and C) and compared, as in \hyperref[fig:Fig5]{Fig.~\ref{fig:Fig5}b}, to the case of statistical independence.  Because neighboring sites tend to have a small, but constant phase difference (cf. \hyperref[fig:Fig5]{Fig.~\ref{fig:ExampleSyncronization}b}) we predict that these probabilities should also be enriched. In contrast to this prediction, we found that methylated CpGs are less likely to be in an adjacent state than expected by statistical independence. 

The reason for this discrepancy is that active turnover of DNA methylation is restricted to accessible and typically active regions of the genome~\cite{parry2021active}.  Because these regions form a small fraction of the entire genome, the analysis conducted so far is statistically dominated by ``passive'' sites not undergoing active turnover. Because this analysis reflects the strong and generic correlations of DNA methylation marks, it serves as a negative control for a deeper-going analysis below.

To specifically test our predictions for genomic regions that undergo active turnover, we filtered for putatively oscillating sites. We defined these sites by the requirement that at the mark 5hmC was detected in at least one of the sequenced cells. For these sites, we also found that nearby sites are more likely to be in the same stage of the cycle compared to the null hypothesis. Significantly, unlike the negative control in \hyperref[fig:Fig5]{Fig.~\ref{fig:Fig5}d}, CpGs in close proximity are simultaneously enriched to be in adjacent stages of the cycle, \hyperref[fig:Fig5]{Fig.~\ref{fig:Fig5}e}. This effect is stronger in serum conditions than in 2i conditions. The relatively weaker effect for cells in 2i conditions is expected from the fact that these cells do not express the Dnmt3 genes such that active turnover of DNA methylation is limited by the relatively inefficient de-novo catalytic activity of DNMT1. Taken together, these empirical findings indicate signatures of phase locking. This phase locking is specific for putatively oscillating sites and, as expected, is not found in the rest of the genome.

\section*{DISCUSSION AND CONCLUSION}

DNA methylation turnover is an active, epigenetic phenomenon that has been associated with the control of key cell fate decisions. Yet, its genomic distribution, temporal evolution, and its functional role remain unknown. This is in part due to limitations in experimental technologies, which puts emphasis on approaches guided by theoretical predictions. In this work, we combined analytical calculations, simulations, and analysis of sequencing data to show that chromatin-mediated interactions between local biochemical cycles can give rise to emergent phase-locking dynamics. We found that phase-locked domains coarsen and that genomic disorder arrests the coarsening dynamics, resulting in the formation of stable phase-locked domains.

We qualitatively tested these results using single-cell sequencing data. We detected qualitative signatures of phase-locking behavior that are specific to regions undergoing DNA methylation turnover. Such tests are currently limited by the constraints of sequencing technologies, which cannot capture dynamic information and have a trade-off between the resolution of the biochemical cycle and the genomic coverage in individual cells.
Our work suggests that understanding the role of active DNA methylation turnover may require novel experimental approaches that capture multiple states of DNA methylation turnover in single cells with high coverage, but that may be limited to small genomic regions, such as enhancers or inducible promoters.

Oscillators are an ubiquitous feature of biological systems. They are used to perform a variety of functions, including timing, as in the cell cycle or the circadian rhythm, and signal processing, as in neurolal oscillators. In somitogenesis, coupled genetic oscillators lead to traveling waves are used to pattern the early embryo. Our work demonstrates the emergence of surprisingly complex patterns arising from DNA methylation turnover. This raises the question whether the functional significance of DNA methylation turnover might lie in these emergent patterns as opposed to the turnover of individual CpGs.
%Genomic disorder is important.

Driven oscillator models have been extensively studied to understand synchronization phenomena. Our theoretical work highlights that chromatin-mediated nonreciprocal interactions give rise to phase locking behavior that differs from models that have been studied so far in this field. It shows that active processes in a chromatin context provide an interesting testing ground for non-equilibrium physics.

%Overall, our findings underscore the importance of a biophysical perspective in understanding epigenetic regulation. DNA methylation turnover can exhibit coordinated dynamics driven by the spatial organization of the genome. The existence of stable phase-locked domains indicate how DNA methylation is regulated on the small scales. Our framework offers a starting point for future experimental and theoretical studies aimed at unraveling the emergent, collective behavior of the epigenome.

\section*{Acknowledgments}
This project has received funding from the European Union’s Horizon 2020 research and innovation programme under the Marie Skłodowska-Curie grant agreement No 101034413. The computations in this paper were run in part on the the FASRC Cannon cluster supported by the FAS Division of Science Research Computing Group at Harvard University and the cluster of the Max-Planck-Institute for the Physics of Complex Systems.  

\section*{Author Contributions}
Conceptualization, supervision and writing: F.O. and S.R.; numerical simulations: F.O. and M.G.; 
theoretical analysis: F.O.; experimental data processing: O.B.; experimental data analysis: F.O.\\
$^*$ rulands@lmu.de

\bibliography{Lib}

\newpage

\pagestyle{empty}
\setcounter{figure}{0} 
\renewcommand{\thefigure}{S\arabic{figure}}

\renewcommand{\appendix}{\par
  \setcounter{section}{0}
  \setcounter{subsection}{0}
  \gdef\thesection{\Alph{section}}
}

\renewcommand{\theequation}{S.\arabic{equation}}
\setcounter{equation}{0}

\clearpage
\appendix

\section*{Supplementary Material}

\section{Discrete phase expansion}
\label{sec:DiscretePhaseExpansion}

In this section, we give details for the Van Kampen expansion of the master equation \eqref{eq:MasterEquationClock}. After expanding the variables with respect to the system size as:  $\phi_i = \Omega \Phi_i(t) + \Omega^{1/2}\xi_i(t)$,  the l.h.s of Eq.~\eqref{eq:MasterEquationClock} is,
\begin{equation}
\sum_i \frac{d P(\vec{\phi},t) }{d \phi_i}= \frac{d\Pi}{dt} - \sum_i\left[   \Omega^{1/2}\frac{d \phi_i}{dt}\frac{d\Pi}{d\xi_i}\right] \, .
\end{equation}
We rewrite the term in the r.h.s of the master equation as
\begin{equation}
\begin{split}
&\sum_i \left[\left(\omega_i + k_{i}(\vec{\phi},\phi_i - 1 \right)\right]P(\vec{\phi},\phi_i - 1)  =\\
&\sum_i\mathrm{E}_i^{-1} \left[\left(\omega_i + k_{i}(\vec{\phi},\phi_i \right)\right]P(\vec{\phi},\phi_i) ,
\end{split}
\end{equation}
where the introduced operator $\mathrm{E}^{\pm}_i$ acts on everything to the right as,
\begin{equation}
\mathrm{E}_i^{\pm} G(\vec{\phi}) = G(\vec{\phi},\phi_i \pm 1) \,.
\end{equation}
$G(\vec{\phi})$ is a general function of the phase and we use the notation  $G(\vec{\phi},\phi_i \pm 1)$ to compactly indicate the function $G\left( \phi_1,\ldots,\phi_i \pm 1, \ldots \phi_N\right)$. The operators $\mathrm{E}^{\pm}_i$ in the system size expansion are approximated to the highest order in $\Omega$ as
\begin{equation}
\mathrm{E}_i^{\pm}  \sim 1 \pm \Omega^{1/2}\frac{\partial}{\partial \xi_i} + \frac{1}{2} \Omega^{-1/2} \frac{\partial^2}{\partial \xi_i^2}.
\end{equation}
Upon using the expansion of the operators and collecting highest order terms in $\Omega$ we get,

\begin{equation}
\begin{split}
   &  \frac{d\Pi}{dt} - \sum_i \left[ \Omega^{1/2}\frac{d \Phi_i}{dt}\frac{\partial \Pi}{\partial \xi_i} \right] = \\
   & \sum_i -w_i\left[\Omega^{1/2}\frac{\partial}{\partial \xi_i} - \frac{1}{2}\frac{\partial^2}{\partial\xi_i^2}\right]\Pi \\
   & -  \Omega^{1/2}\frac{\partial}{\partial \xi_i}\left[k(\Phi_i) + \Omega^{-1/2}\frac{\partial k}{\partial \Phi_i}\xi_i\right]\Pi \,.
\end{split}
\end{equation}
Collecting terms in power of $\Omega^{1/2}$, and using the chain rule we get
\begin{equation}
   \frac{d\Phi_i}{dt} = w_{i} + f_1(\Phi_i)\sum _{k=1}^{N}\frac{J_1 e^{{{-m|k -i|}}}}{|k-i|^{\lambda}}f_2(\Phi_k) ,  
\end{equation}
which is the mean field equation and where we simply made $k(\Phi)$ explicit. The next lowest order in $\Omega$ expresses the dynamics of fluctuations as a Fokker-Planck equation,
\begin{equation}
  \frac{\partial \Pi}{\partial t} = \sum_i \left[ (w_i + k(\Phi))\frac{\partial^2 \Pi}{\partial \xi_i^2} - w_{L}\frac{\partial k(\Phi)}{\partial \Phi_i}\frac{\partial (\xi_i \Pi_i)}{\partial \xi_i}\right] \,.
\end{equation}
For small values of $k(\Phi)$ the noise is dominated by $\omega_i$ and it is a Gaussian white noise arriving to   Eq.~\eqref{eq:langevinOscillators}.

\section{Derivation of the continuity equation}
\label{sec:FPDerivation}

In order to quantify the degree of system-wide synchronization we study the behavior of the Kuramoto order parameter $r(t)$ \cite{Kuramoto2003} by,
\begin{equation}
    r(t)e^{i\psi(t)}= \frac{1}{N}\sum_{i=1}^N e^{i\phi_i(t)} \, .
\end{equation}
In order to find an analytical expression for $r(t)$ we first define the generator of moments \cite{Perez1997,Bonilla1998,Acebron2005},
\begin{equation}
\label{eq:moments}
H^c_{k,q} = \frac{1}{N}\sum_{j=1}^{N}\overline{\langle e^{ik\phi_j} e^{iqj} \rangle w_j^{c}} \, ,
\end{equation}
where $\overline{\left( .\right)}$ is the average over the distribution $g(\omega)$ of intrinsic frequency $\omega_i$ and $\langle \left(.\right)\rangle$ is an average over all possible realizations of noise.
In  equation \eqref{eq:moments}, we employed a Fourier transform of the fields. Hence, $k$ is dual to $\phi$ and $q$ is dual to the position along the DNA sequence.

In Eq.~\eqref{eq:moments} we introduced the moment generating function for $\omega_i$ with respect to both stochastic noise and intrinsic noise given by the possible non-delta distribution of intrinsic frequencies. In It\^{o} convention, the time evolution of an  arbitrary function $F(.)$ of a given stochastic process $\phi_i$ ( $i=1,...,N$) and with noise amplitude $\sqrt{2D_i}$ is,
\begin{equation}
\label{eq:Ito}
\partial_t F(\vec{\phi}) = \sum_j \left[\partial_{\phi_j}F(\phi) \partial_t \phi_j + \sum_k \frac{\partial^2 F(\vec{\phi})}{\partial \phi_j \partial \phi_k} \sqrt{D_j D_k}\right] \, .
\end{equation}
As the previous equation is valid for any function $F$, we compute the function $ \partial_t \langle e^{ik\phi_j}e^{iqj} \rangle$. We substitute in Eq.~\eqref{eq:Ito} the stochastic process in which trajectory is given by Eq.~\eqref{eq:langevinOscillators} hence obtaining (after using the standard property of stochastic calculus)
\begin{equation}
\frac{\partial \langle e^{ik\phi_j}e^{iqj}\rangle }{\partial t} =  ike^{ik\phi_j}e^{iqj}G(\phi_j) - k^2 e^{ik\phi_j}e^{iqj}D_j \,.
\end{equation}
We then multiply the previous equation by $\omega_j^c$, sum over all $j =1,...N$, divide by $N$ and average over the frequency distribution,
\begin{equation}
\label{eq:MomentsKuram}
\begin{split}
   & \frac{1}{N}\sum_j
\overline{\frac{\partial \langle e^{ik\phi_j}e^{iqj}\rangle \omega_j^c}{\partial t}} = \\
& \frac{1}{N}\sum_j \overline{\omega_j^c \left[ike^{ik\phi_j}e^{iqj}G(\phi_j) - k^2 e^{ik\phi_j}e^{iqj}D_j\right]},
\end{split}
\end{equation}
 With $G(\phi_j) = \omega_j+ f_1(\phi_j)\sum _{w=1}^{N}\frac{J e^{{{-m|w -j|}}}}{|w-j|^{\lambda}}f_2(\phi_w) $. The l.h.s of the previous equation is simply $\partial_t H^m_{k,q}$ as defined in Eq.~\eqref{eq:moments}. To simplify the r.h.s, we introduce the Fourier representation of $f_1$, $f_2$ as
\begin{equation}
\label{eq:FT}
\begin{split}
f_1(\phi_j) &= \sum_{n = -\infty}^{\infty}a_n e^{i n\phi_j}\\
f_2(\phi_w) &= \sum_{l = -\infty}^{\infty} b_l  e^{i l\phi_w}
\end{split}
\end{equation}
Being $\frac{ e^{{{-m|w -j|}}}}{|w-j|^{\lambda}}$ just a function of the difference we can define its Fourier transform as 
\begin{equation}
    \frac{ e^{{{-m|w -j|}}}}{|w-j|^{\lambda}} = \sum_{s = -\infty}^{\infty}r_s e^{i s (w-j)} \,.
\end{equation}
The first term on the r.h.s of Eq.~\eqref{eq:MomentsKuram} is given by
\begin{equation}
  \frac{J}{N} \sum_j\sum_{n,l,s} \sum_{c}a_n b_l r_s ike^{i(k+n)\phi_j}e^{i(q-s)j}  e^{il\phi_w}e^{isw} \omega_j^c \,.
\end{equation}
As $\frac{1}{N} \sum_w e^{il\phi_w}e^{isw} = H_{l,s}^0$, the previous equation  simplifies to
\begin{equation}
(ik)J N \sum_{n,l,s} a_n b_l r_s H_{k+n, q-s}^c H_{l, s}^0  \,.
\end{equation}
The other terms can be computed in a similar way and the resulting dynamical equation for the moments is
\begin{equation}
\label{eq:momentsdynamics}
\begin{split}
  &  \partial_t H_{k,q}^c =(ik)J N \sum_{n,l,s} a_n b_l r_s H_{k+n, q-s}^m H_{l, s}^0 +\\
    & (ik)H_{k,q}^{c+1} - k^2H_{k,q}^{c+1} \,.
\end{split}
\end{equation}
To obtain a more compact expression  we define the  generating function
\begin{equation}
    \chi(\theta,y,z,t) = \sum_{k = -\infty}^{\infty}\sum_{c = \infty}^{\infty}\sum_{q = -\infty}^{\infty} e^{-ik\theta} e^{-iqz} \frac{y^c}{2\pi c!} H_{k,q}^{c} \,,
\end{equation}
and we show how all the terms of Eq.~\eqref{eq:momentsdynamics} can be rewritten in terms of this function. We analyze, at first, the second term on the r.h.s of  \eqref{eq:momentsdynamics}, which can is simplified to, (upon summing over $m,k,q$ and multiplying by $ \frac{y^c}{2\pi c!} $) 
\begin{equation}
- \frac{\partial}{\partial \theta}  \sum_{k ,c,q} e^{-ik\theta} e^{-iqz}\frac{\partial}{\partial y}  \frac{y^{c+1}}{2\pi c!} H_{k,q}^{c+1}\,.
\end{equation}
In terms of the function, $\chi$, the previous equation is,
\begin{equation}
-\frac{\partial^2}{\partial \theta \partial y} \chi(\theta,y,z,t) \,.
\end{equation}
Following the same procedure, the third term on the r.h.s of Eq.~\eqref{eq:momentsdynamics} is rewritten as
\begin{equation}
 \frac{1}{2}  \frac{\partial}{\partial y}  \frac{\partial}{\partial \theta}D(y) \frac{\partial}{\partial \theta} \chi(\theta,y,z,t) \,.
\end{equation}
We proceed now to write the terms describing interactions in terms of the moments. Initially, we write it as
\begin{equation}
    -JN \frac{\partial }{\partial \theta}\left[ \sum_{n,l,s}a_nb_lr_se^{inx}e^{isz} \chi(\theta,y,z,t) H_{l, s}^0\right] \,,
\end{equation}
where we multiplied and divide by $e^{inx}e^{isz}$  so that $H_{k,q} \rightarrow H_{k+n,q-s}$. We further define a term
\begin{equation}
    \nu(\theta,y=0,z,t) =J N  \left[ \sum_{n,l,s}a_nb_lr_se^{inx}e^{isz} H_{l, s}^0\right] \,
\end{equation}
and writing Eq.~\eqref{eq:FT} inversely and going back to the space of function defined on $\phi$ ($\theta$ in this notation) we obtain,
\begin{equation}
    \nu(\theta,y=0,z,t) =J N\left[ f_1(\theta)\sum_{l,s}b_lr_se^{isz} H_{l, s}^0\right] \,.
\end{equation}
We explicit all the other terms in the same way as,
\begin{equation}
      \nu = J \left[ f_1(\theta)\int d\hat{\theta} \int d\hat{z}\frac{ e^{-m \hat{z}}}{|\hat{z}|^{\lambda}}  f_2(\hat{\theta}) \chi(\hat{\theta},y=0,z-\hat{z},t)\right] \,,
\end{equation}
wtih $\, z \in (0,1)$. Altogether the time evolution of the moment generator $\chi$ is given by,
\begin{equation}
  \label{eq:generalcontinuity}
\begin{split}
\partial_t \chi(\theta,y,z,t) &= -\frac{\partial}{\partial \theta}[\nu(\theta,y,z,t)\chi] + \\
&\frac{\partial}{\partial y} \frac{\partial}{\partial \theta}D(y)\frac{\partial}{\partial \theta}\chi(\theta,y,z,t) - \frac{\partial \chi}{\partial \theta \partial y} \, ,  
\end{split}
\end{equation}
where the drift term takes the form
\begin{equation}
\nu = y + J \left[ f_1(\theta)\int d\hat{\theta} d\hat{z}\frac{ e^{-m \hat{z}}}{|\hat{z}|^{\lambda}} f_2(\hat{\theta}) \chi(\hat{x},0,z-\hat{z},t)\right]
\end{equation}
with diffusion term reads $D(y)  = 2y$. 
Defining $\rho(\theta,\omega,z,t)$ as the density of oscillators with phase $\theta$, position $z$ and frequency $\omega$ at time $t$. The  continuity equation is  given by exploiting the relationship $\chi(\theta,y,z,t) = \int d\omega g(\omega) e^{y\omega} \rho(\theta,\omega,z,t)$ and setting the diffusion to zero for the deterministic limit. 

\section{Phase-locked solution}
\label{sec:StatSol}

Equation~\eqref{eq:continuity} does not admit asynchronous stationary solutions. The control functions $f_1, f_2$ break the rotational symmetry such that the equations are not invariant under a shift of the fields (oscillators). Indeed, the first integral in Eq.~\eqref{eq:continuity}  will never vanish for an asynchronous solution as long as $\int \text{d}\theta f_2(\theta) \neq 0$. We also have to consider  consistency with the biological reality: the distribution of frequencies for each oscillator, $g(\omega)$ has positive support as, on average, the biochemical cycle advances in the positive direction. Typically one can shift the distribution by moving to a rotating reference frame with an angular velocity given by the median of $g(\omega)$  \cite{Acebron2005}. However, in our model, due to the lack of rotational symmetry, this transformation does not simplify the analytical calculations. For general functions $f_1(\theta_i) f_2(\theta_j)$, it is sometimes possible to change to a coordinate system that allows the asynchronous solution to be the stable \cite{Daido1992}, as long as the product $f_1(\theta_i) f_2(\theta_j)$  can be written as a $2\pi$-periodic function of the phase difference $\theta_i -\theta_j$, with $g$ being a function that is $2\pi$ periodic and well defined. As the $f_{1,2}(\theta)$  are step functions, it is not always possible to write the product in the rotationaly symmetric form. Moreover, since the asynchronous solution is never stable, a stability analysis as in \cite{Strogatz1992,Strogatz2000} or a power series expansion \cite{Ott2008} around this state does not give any useful insight for this process; the stationary states are non-trivial even for a vanishing strength of interactions $J$.  We thus look for an alternative order parameter, which still remains experimentally accessible. A natural order parameter is the fraction of sites for which $f_2(\phi)=1$, namely $m$. Biologically, this is the average DNA methylation. For the biological control functions, $m = 1/2$ in a phase where the oscillators are uncoupled and it is greater than $1/2$  whenever $J>J_c$.  We are now interested in finding the possible behaviour of the collective dynamics of oscillators. To this end, we search for a stationary solution of Eq.~\eqref{eq:continuity} 

\section{Stationary solution}
\label{sec:Mvalue}

A spatially homogeneous stationary solution of Eq.~\eqref{eq:continuity} is ($\omega \geq 0$ in the model definition),
\begin{equation}
\label{eq:stationarysolutions}
\rho =
       \frac{C(\omega)}{|\omega + H(\pi/2 -\theta) A(\lambda) J|} \, ,
\end{equation}
where
\begin{equation}
    A(\lambda)=\left(\pi \int \upd \omega\, \frac{g(\omega) C(\omega) }{\omega} \right)^{\lambda} \int_0^{ \pi \int d\omega \frac{g(\omega) C(\omega)} {\omega}} \text{d}y\, \frac{e^{-|y|}}{|y|^{\lambda}}\, ,
\end{equation}
and $\Gamma(1-\lambda,0,m) = \int_0^{m}\text{d}y \,e^{-|y|}/|y|^{\lambda}$ is the generalised incomplete gamma function. $H(x)$ denotes the Heaviside step function. $C(\omega)$ is set by the normalization of the probability density function $\rho$, \cite{Acebron2005}  ($\int_0^{2\pi} d\theta \rho = 1$),  
\begin{equation}
    C(\omega) = \frac{2|\omega|| A(\lambda) J+ \omega |}{\pi  \left(|\omega| + 3 |A(\lambda)
   J|\right)}\, .
\end{equation}
To understand how the degree of synchronization changes as a function of the coupling strength, $J$, we compute  $m$  from Eq.~\eqref{eq:definition_meth} using  the stationary distribution of $\rho$, Eq.~\eqref{eq:stationarysolutions}.
\begin{equation}
\label{eq:mtemp}
    m =   \int_0^{\infty} \upd\omega  \frac{  2(A(\lambda) J+\omega)}{3A(\lambda)
   J+4 \omega }  g(\omega) \, .
\end{equation}
where$ A(\lambda) = A(\lambda) =  m^{\lambda} \Gamma(1-\lambda,0,m )$.

\section{Numerical Simulation Methods}
\label{sec:NumericalMethods}
All mean-field simulations are run with a 4th order Runge-Kutta method with discrete time-step of $dt = 10^{-3}$ and $L= 10^3$ lattice sites, of which the first $200$ are shown. The exponent of the long-range kernel, Eq.~\eqref{eq:KernelDefinition} is $\lambda = 1/3$. 
Domains boundaries in \hyperref[fig:Fig4]{Figure.~\ref{fig:Fig4}a} are defined whenever two neighboring sites have a phase difference greater than  $ dt \, max_i(\omega_i + k_i)$

For the stochastic simulations of Eq.~\eqref{eq:MasterEquationClock}, we used the Gillespie algorithm~\cite{gillespie1977exact}. Each point on the lattice, set equidistant from each other, mimics a single CpG site, with the genome size being a 1000 sites under periodic boundary conditions. Each site $i \in [1,2,...1000]$ is defined by its position  along the genome $i$, its clock state $CS_i$ and an intrinsic rate parameter $\omega_i$ drawn from an exponential distribution with the unitary mean. We tested simulation with the number of clock states, $N_{states}$, ranging between 10 and 100 with qualitatively consistent results.%  $N_{\text{states}}= 100$ states such that $CS_i \in [1,2,...100]$. 
We also tested The clock state on each site was initialized randomly from a uniform distribution.

%For each timestep of the simulation, a single site in the genome is chosen to increase by one in the clock $(CS_i \rightarrow CS_i + 1)$. We chose a site for the update step as following: the rates per site are summed up across the genome, normalised by number of sites, and added to the cumulative sum of rates per site. We then sample from the cumulative distribution and pick a site, as in \cite{gillespie1977exact}. As we parse the genome sites, the first site that has a cumulative rate higher than a randomly chosen number is updated to the next state in the clock. 
%We measured the average methylation of all sites at each time step $\delta_t$. 
We defined the control function $\psi_2$ as between $\frac{\pi}{2} \rightarrow 25$ and $\frac{3\pi}{2} \rightarrow 75$. The value of $m$ was defined as the fraction of sites in phase interval $\psi_2$. %This was then averaged over number of states in the clock (100) as well as how many sites contributed to the sum (variable in each time step). 
We also calculated the Kuramoto order parameter for each time step defined as $r = \left[(\sum_i \cos(2 \pi CS_i)\right]^2 + \left[(\sum_i \sin(2 \pi CS_i))^2\right]^{1/2}$, where $CS_i\in\{1,2,\ldots,N_{states}\}$./
%In each time step, the rates of each site on the genome are also updated. If a given site on the genome has a clock state greater than $\frac{\pi}{2} \rightarrow 25$, the rate of that site remains the intrinsic $\omega_i$ since that site is not interacting. If it is in the initial interval  $\psi_1 \in [0,\frac{\pi}{2} \rightarrow 25]$ states and can interact, we update it's rate in accordance to our power law. We then find the nearest site with a clock state in $\psi_2$, between $\frac{\pi}{2} \rightarrow 25$ and $\frac{3\pi}{2} \rightarrow 75$ to mimic the second phase. The distance is calculated to the right $(D_r)$ and to the left $(D_l)$ of each site in the first phase. Such sites from the first phase get an updated rate of $\omega_i + J ((\frac{1}{D_r})^{\lambda} + (\frac{1}{D_l})^{\lambda})$. Each simulation is run with a constant value of J, or coupling strength. We run multiple simulations, for $J$ ranging from $0$ to $10$. $\lambda$ is in accordance with experiments \cite{olmeda2021inference} and it is $\lambda = \frac{1}{3}$.
Each simulation ran for a maximum time of $T=L^2/N_{\text{states}}$ and we measured the average methylation once every 100 time steps.

\section{Data Processing and Analysis}
\label{sec:DataAnalysis}
Raw sequencing used in this study is available from the Gene Expression Omnibus (\href{https://www.ncbi.xyz/geo/}{GEO}) with accession number \href{https://www.ncbi.xyz/geo/query/acc.cgi?acc=GSM5929312}{GSE197740} and sample number \href{https://www.ncbi.xyz/geo/query/acc.cgi?acc=GSM5929312}{GSM5929312} (embryonic stem cells, mouse). The bioinformatics analysis is carried out as described in Ref.~\cite{experimentaldatabai} with a slight modification. Briefly, we aligned the sequencing data to the mouse (GRCh38) genome reference and used default parameters for both alignment and quality control using a custom pipeline. Upon extracting methylation and hydroxymethylation states of cytosines in CpG contexts, we did not perform any binning. The pipeline used for the bioinformatics analysis is available in Ref.~\cite{bektas2024simpleseq}.
In \hyperref[fig:Fig5]{Figure~\ref{fig:Fig5}b,c} we use the data in bulk with the only restriction to have sites that were tested for both 5mC and 5hmC. In \hyperref[fig:Fig5]{Figure~\ref{fig:Fig5}b,c} we subset the data to sites where at least one site in all cells has positive reads for 5hmC and negative reads for 5mC.

%\bibliography{Lib}

\end{document}